\documentclass{raa}       
\usepackage{graphicx,times}           
\usepackage{natbib}
\usepackage{amssymb,amsmath}
\bibpunct{(}{)}{;}{a}{}{,}
\usepackage{float}
\usepackage[pagebackref=true]{hyperref}
\hypersetup{colorlinks = true, linkcolor = green, anchorcolor = red, citecolor = blue, filecolor = red,  urlcolor = red}

\begin{document}
   \title{Spotless days and geomagnetic index as the predictors of solar cycle 25}
   \volnopage{Vol.0 (20xx) No.0, 000--000}      
   \setcounter{page}{1}          
      \author{Dipali S. Burud
       \inst{1,9}
      \and Rajmal Jain
       \inst{2}
      \and Arun K. Awasthi
      \inst{3,4}
       \and Sneha Chaudhari
      \inst{9}
       \and Sushanta C. Tripathy
      \inst{5}
       \and N. Gopalswamy
      \inst{6}
       \and Pramod Chamadia
      \inst{7}
       \and Subhash C. Kaushik
      \inst{8}
      \and Rajiv Vhatkar
      \inst{1}
   }

   \institute{Department of Physics, Shivaji University, Kolhapur, India\\
        \and
             Physical Research Laboratory, Navrangpura, 380009 Ahmedabad; {\it profrajmaljain9@gmail.com}\\
        \and
              CAS Key Laboratory of  Geospace Environment, Department of Geophysics and Planetary Sciences, University of Science and Technology of China, Hefei 230026, China\\
               \and 
             CAS Key Laboratory of Solar Activity, National Astronomical Observatories, Beijing 100101, China\\ 
         \and
            National Solar Observatory, 3665 Discovery Drive, Boulder, CO 80303, United States\\
         \and
             Goddard Space Flight Centre, NASA, Washington, USA\\
         \and
            Department of Physics, Govt. P. G. College, Satna, M. P., India,\\
         \and
            Department of Physics, Govt. PG Autonomous College, Datiya, M. P, India\\
         \and
             Sh. M. M. Patel Institute of Sciences and Research,
                         Kadi Sarva Vishwavidyalaya, Gandhinagar, India\\
\vs\no
   {\small Received~~20xx month day; accepted~~20xx~~month day}}

\abstract{We study the sunspot activity in relation to spotless days (SLDs) during the descending phase of solar cycle $11$--$24$  to predict the amplitude of sunspot cycle $25$. For this purpose, in addition to SLD, we also use the geomagnetic activity (aa index) during the descending phase of a given cycle. A very strong correlation of the SLD (R=$0.68$) and aa index (R=$0.86$) during the descending phase of a given cycle  with the maximum amplitude of next solar cycle  has been estimated.The empirical relationship  led us to deduce the amplitude of cycle $25$ to be  99.13$\pm$ 14.97  and 104.23$\pm$ 17.35  using SLD and aa index, respectively as predictors.Both the predictors provide comparable amplitude for solar cycle $25$ and reveal that the solar cycle $25$ will
be weaker than cycle $24$. Further we derive that the maximum of cycle $25$  is likely to occur between  February and March 2024. While the aa index has been used extensively in the past, this work establishes SLDs as another potential candidate for predicting the characteristics of the next cycle.
}

   \authorrunning{Burud, Jain, Awasthi et al.}            
   \titlerunning{{Spotless days and geomagnetic index probing solar cycle 25 } }  

   \maketitle

%
%
\section{Introduction}           
\label{sect:intro}
The magnetic field is generated in the solar interior through the dynamo mechanism via the motion of plasma and seen as a dark spot with strong magnetic field. These observed sunspots are varying in number between $0$ (lowest) and $250$ (highest) in approximately $11$ years period well known as Schwabe cycle. The solar cycles are characterized by their length, amplitude, rise time and fall time etc. The solar activities such as the sunspot number, sunspot areas, total solar irradiance,Coronal Mass Ejection (CME), solar flares, facular area and $10.7$ cm solar radio flux ( $F10.7$) etc, show cyclic behavior as observed by the various researcher \citep{Jain1986, Jain1997, Krivova2002, Braun2005, Atac2006, Kilcik2010, Kilcik2012, Hathaway2002, Javaraiah2019, ElBorie2020}.  The period of Schwabe cycle varies between $7.4$ and $14.8$; with the ascending and descending phase of $4$-$5$ and $5$-$6$ years, respectively \citep{Usoskin2003}.These types of periodicity are very useful in understanding the solar phenomena which has the long-term change in climate and short-term changes in space weather \citep{Nandy2007}.

In a given solar cycle, there are many days without sunspots, which are known as spotless days (SLDs).   The number of SLDs in a year or in a given cycle can be obtained from the web site \url{(www.sidc.be/silso/datafiles)} which covers the period $01$ January $1818$ to December $2020$. The SLDs are generally noted during the descending phase and reach to maximum in the sunspot minimum year.  SLDs are also found to be present during the maximum year. The highest number of SLDs ($311$) occurred in the year $1913$ of cycle $14$.The maximum number of SLDs ($973$) in a given cycle was found in cycle $12$. The total numbers of SLD in cycle $23$ (August $1996$ to $31$  December $2019$) are found to be $573$, similar to cycle $15$ ($556$) and cycle $16$ ($541$). However  the cycle $24$, showed a total $913$  SLDs untilDecember $2019$ (Burud et al. $2021$, in preparation).

The highest value of SLD in a given cycle has been used to define the minimum phase of the corresponding cycle by \citep{Waldmeier1961} and \citep{McKinnon1987}.  The  authors suggested that the date of sunspot minimum can be decided by the number of SLD. \citep{Wilson1995} used the timing of occurrence of the first observed SLD in a given cycle with various other timings such as  time period between the solar minimum to the first SLD observation after the solar maximum, time period between the solar maximum to the first SLD observation after the solar maximum etc. to predict the minimum period of solar cycle $22$. Similar result has been observed by \citep{Hamid2006} and \citep{Carrasco2016}.  \citep{Harvey1999} showed that the SLD is not the only  parameter to decide the minimum of a solar cycle but other parameters such as the monthly averaged sunspot number, the number of regions (total, new- and old-cycle) etc. need to be considered. \citep{Helal2013} indicates that the relation between the SLDs along two years intervals around the preceding minimum can be used to predict the amplitude of the next  solar cycle and obtained  a value of $118.2$ for cycle $25$.

 Further, the geomagnetic activity (aa  index) during the descending phase of a given sunspot cycle provides an asymptomatic prediction of the magnitude of annual mean sunspot number of the next sunspot cycle \citep{Jain1997}. Employing aa index as predictor, \citep{Jain1997} estimated amplitude of the solar cycle $23$ to be $166.2$, which is found in agreement with the observed value of  $173.4$. However, it may be noted that \citep{Jain1997} extended the period of descending phase to five years (minimum plus previous four years) to predict the amplitude and timing of the next solar cycle as predictor. Further, it would  be interesting to explore the relationship between the two predictors:   SLD and aa index.

In this paper, we carry out a statistical analysis of SLD and aa index during descending phase of the solar cycle $11$ to $24$, to investigate the characteristics of the solar cycle $25$. This paper is organized as follows: in section $2$ we give sources and description of the data, analysis and results are presented in section $3$ and discussions and brief conclusions are presented in section $4$ and $5$ respectively.


\section{Sources of Data}
\label{sect:Obs}
 We consider  the sunspot number (R) that has recently been published and obtain from SILSO web site to study the physical properties of SLD from solar cycle $06$ to $24$ \citep{Wilson1996, Li2005}. The SLD data are taken for the period January, $1868$ to December $2019$ covering cycle $11$ to $24$ since the aa index is available for the  period. In order to minimize the inconsistency in calendar years  related to sunspots numbers \citep[see][]{Petrovay2020} a $13$-month  boxcar averages of monthly mean sunspot numbers is calculated as follows:
\begin{equation*}
  R\textsubscript{m}=\frac{1}{24}(R\textsubscript{i,-6}+2 \sum_{j=-5}^{j=5} R\textsubscript{i,j}+R\textsubscript{i,6})
\end{equation*}
				
where, $R\textsubscript{m}$ is the $13$-month smoothed maximum annual mean sunspot number which is calculated by using the mean of monthly sunspot numbers (R\textsubscript{j}) over $13$ months centered on the corresponding month (R\textsubscript{i}).

The aa index has been taken from British geological survey where the data is available from $1868$ at \url{(www.geomag.bgs.ac.uk/data_service/data/magnetic_indices/aaindex)}. The aa index provides a daily average level of the geomagnetic activity measured in nT. Eight datasets of $3$-hourly values of aa index are averaged to obtain the daily aa index for a given day. As mentioned earlier, the aa index is taken for the period January, $1868$ to December $2019$, representing cycle $11$ to $24$.

\section{Analysis and results}
\label{sect:analysis}
\subsection{Prediction of the 13-month smoothed maximum annual mean sunspot number}
The SLD data is taken from website maintained by Solar Influences Data Center ( SIDC), Royal observatory of Belgium. It is also observed that the SLD can occur between two days when the sunspots have been located on the solar disk. Generally, during the descending phase of the solar cycle,  we have observed that there are many continuous days without sunspots. We define such continuous  period of days observed as consecutive spotless days. Our data of SLD show that the consecutive days  between $1$ to $92$ days, as presented in Table $1$. This table allows us to study occurrence frequency of SLD distribution with respect to number of consecutive days. As shown in figure $1$, we find the distribution follows a negative exponential function.

\begin{table}
\centering
\caption{Occurrence frequency of SLD for descending phase of solar cycle $11$-$24$}
\label{tab1}
\setlength{\tabcolsep}{1pt}
\small
 \begin{tabular}{|p{1.7cm}|p{1.7cm}|p{1.7cm}|p{1.7cm}|p{1.7cm}|p{1.7cm}|}
  \hline
\textbf{Consecutive
Days}&\textbf{Occurrence frequency}&\textbf{Consecutive
Days}&\textbf{Occurrence frequency}&\textbf{Consecutive
Days}&\textbf{Occurrence frequency}\\
  \hline
1	&264	&16	&9	&31	&4\\
\hline
2	&140	&17	&7	&36	&2\\
\hline
3	&113	&18	&7	&37	&2\\
\hline
4	&86	&19	&9	&39	&1\\
\hline
5	&60	&20	&7	&40	&2\\
\hline
6	&50	&21	&6	&42	&1\\
\hline
7	&40	&22	&4	&43	&1\\
\hline
8	&38	&23	&2	&44	&1\\
\hline
9	&31	&24	&5	&45	&1\\
\hline
10	&25	&25	&3	&47	&1\\
\hline
11	&17	&26	&6	&49	&1\\
\hline
12	&17	&27	&6	&69	&1\\
\hline
13	&22	&28	&1	&92	&1\\
\hline
14	&15	&29	&1	&	&\\
\hline
15	&20	&30	&3       &        &\\
  \hline
\end{tabular}
\end{table}

\begin{figure}
   \centering
   \includegraphics[width=0.6\textwidth, angle=0]{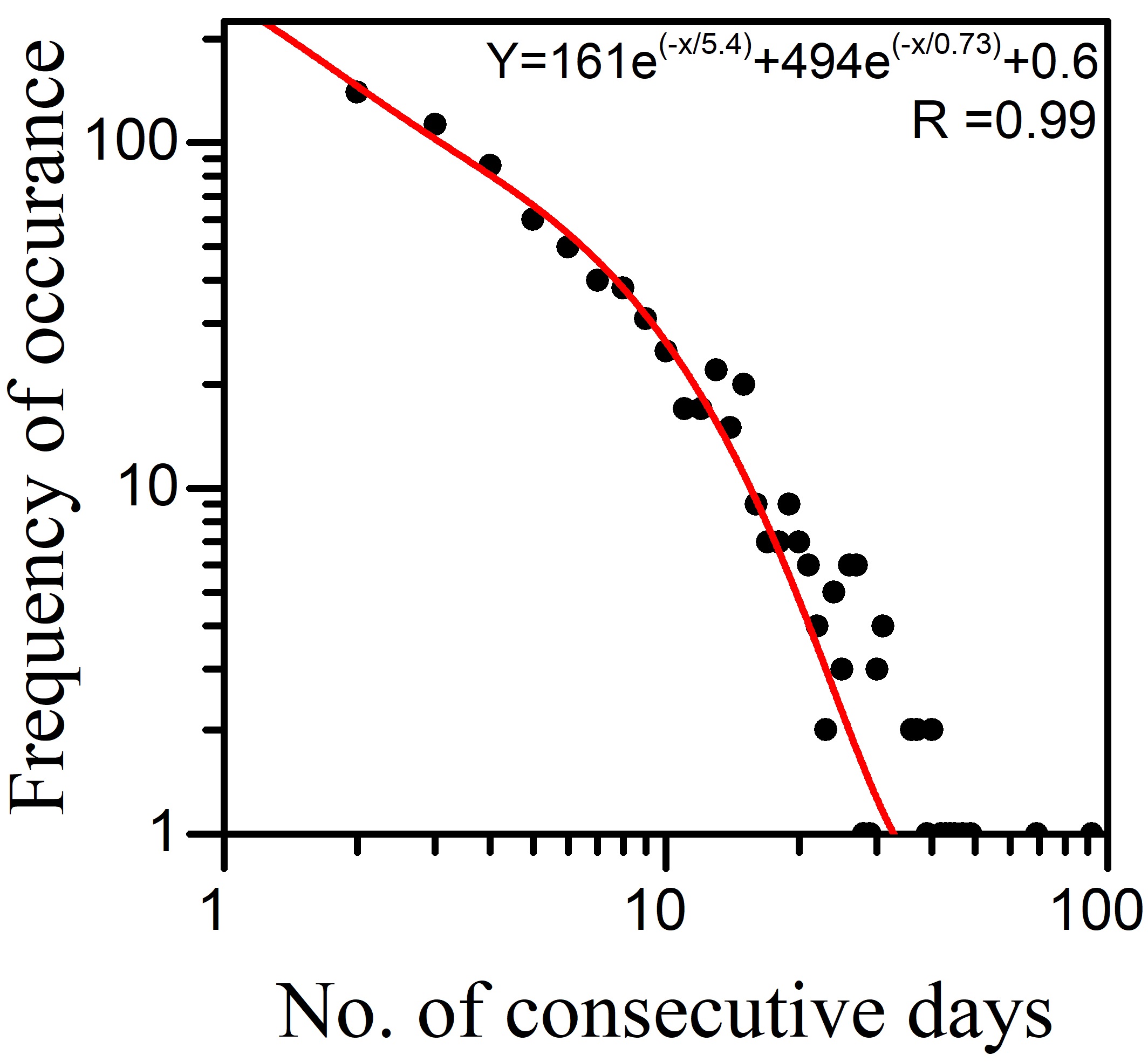}
\caption{Occurrence frequency of SLD as a function of number of consecutive SLD during the descending phase of solar cycle 11 to 24.}
\label{fig1}
\end{figure}

The table $2$  represents the yearwise  comparison of parameters such as, 13 month smoothed annual mean sunspot number ( R\textsubscript{m}), number of  SLDs (SLD) and sum of aa index $(\sum{aa})$ over the desending phases of  cycle $21$, $22$,  $23$ and $24$. 

The Table $2$ reveals that the  total numbers of SLD ($\sum{SLD}$) of the descending phase of cycle $22$ is $1.06$ times  higher relative to cycle $21$, and, similarly cycle $23$ is 2.07 times larger relative to cycle $22$. This trend of incrmeant in the value of SLD has also been observed in the current cycle $24$, in which the SLD of its descending phase  is 1.18 times higher relative to cycle 23. This indicates the  upward trend  in the number of SLD while,13-month smoothed monthly sunspot number (R\textsubscript{m}) and sum of aa index $(\sum{aa})$ shows downward trend  for the long term cycle has begun from cycle 21.This suggests that  Sun is heading towards modern minimum of Gleissberg cycle  \citep{ feynman1990}.
\begin{table}
\centering
\caption{Yearwise comparison of  R\textsubscript{m}, no. of  SLD and sum of aa index $(\sum{aa})$ over descending phase of solar cycle $21$,$22$, $23$ and $24$}
\label{tab2}
\setlength{\tabcolsep}{1pt}
\small
\begin{tabular}{|p{0.72cm}|p{0.85cm}|p{0.64cm}|p{0.77cm}|p{0.72cm}|p{0.85cm}|p{0.64cm}|p{0.77cm}|p{0.72cm}|p{0.85cm}|p{0.64cm}|p{0.77cm}|p{0.72cm}|p{0.85cm}|p{0.64cm}|p{0.77cm}|}
\hline
\multicolumn{4}{|c|}{cycle $21$}&\multicolumn{4}{|c|}{cycle $22$}&\multicolumn{4}{|c|}{cycle $23$}&\multicolumn{4}{|c|}{cycle $24$}\\
\hline
Year&R\textsubscript{m}&SLD&$\sum{aa}$&Year&R\textsubscript{m}&SLD&$\sum{aa}$&Year&R\textsubscript{m}&SLD&$\sum{aa}$&Year&R\textsubscript{m}&SLD&$\sum{aa}$\\
\hline

1982	&114.28	&0	&98782&1992	&93.85	&0	&79691	&2004	&41.90	&3	&67475	&2015	&72.63	&0	&64945\\
\hline
1983	&74.68	&4	&86134&1993	&55.56	&0	&74251	&2005	&28.93	&13	&67653	&2016	&41.51	&27	&58315\\
\hline
1984	&42.20	&13	&84635&1994	&30.23	&19	&85515	&2006	&16.06	&65	&47211	&2017	&21.33	&96	&56635\\
\hline
1985	&17.88	&83	&65833	&1995	&17.27	&61	&63906	&2007	&7.98	&163	&43628	&2018	&8.68	&208	&40438\\
\hline
1986	&13.77	&129	&61308	&1996	&9.10	&165	&54244	&2008	&2.86	&265	&41277	&2019	&3.52	&273	&36404\\
\hline
\end{tabular}
\end{table}

Further in this study, following the technique proposed in \citep{Jain1997}, \citep{Wang2009}, and \citep{Bhatt2009}, we define the descending phase of a given n\textsuperscript{th} cycle to be spanning over a period of five years, which comprises of the minimum year and preceding four years.Subsequently, representative aa index and SLD values of the n\textsuperscript{th} cycle have been estimated by integrating over the entire descending phase duration of five years due to the following reason. Based on the examination of the polarity of the active regions and identification of the location of the ephemeral regions, \citep{Harvey1992} proposed that the overlap between the two cycle is five years. Further, based on a statistical investigation of the time-series observations of $128$ years, \citep{Jain1997} found that the aa index averaged over the period of five years is the effective precursor for estimating the characteristics of the next solar cycle. A total number of aa index $( \sum aa\textsubscript{dsc}) $ and SLDs $( \sum SLD\textsubscript{dsc} $ ), integrated over the descending phase of the n\textsuperscript{th} cycle for the cycle $11$ to $24$ is given in table $3$.  

\begin{table}
\centering
\caption{{Parameters of solar cycle $11$-$24$}}
\label{tab 3}
\setlength{\tabcolsep}{1pt}
\small
\begin{tabular}{|p{0.90cm}|p{2cm}|p{1.80cm}|p{1.2cm}|p{1.5cm}|p{1.5cm}|p{1.68cm}|p{1.68cm}|}
\hline
\textbf{Cycle no.}&\textbf{Descending years}& \textbf{$(\sum{SLD}$)\textsubscript{dsc}} 
&\textbf{$\sum{SLD}$}& \textbf{$(\sum{aa}$)\textsubscript{dsc}} &\multicolumn{3}{c|}{\textbf{R\textsubscript{max}}}\\
\cline{6-8}
 &&&&&\textbf{Observed}&\textbf{Calculated by SLD
                               (error in \%)}&\textbf{Calculated by aa
                               (error in\%)}\\
\hline
11	&1874-1878	&753	&806	&151672	&140&-	&-\\
\hline
12	&1886-1890	&699	&973	&221763	&72.5&85.42(17.8)&	67.8(6)\\
\hline
13	&1898-1902	&845	&909	&141825	&87.9&90.39(2.8)&	91.7(4)\\
\hline
14	&1909-1913	&846	&903	&199465	&64.2&76.96(20)	&64.5(0.5)\\
\hline
15	&1919-1923	&387	&556	&250574	&105.4&76.86(27)&	84.1(20)\\
\hline
16	&1929-1933	&394	&541	&293386	&78.1&119.09(52)	&101.5(30)\\
\hline
17	&1940-1944	&253	&427	&333868	&119.2&118.45(0.6)	&116.06(2.6)\\
\hline
18	&1950-1954	&398	&414	&352692	&151.7&131.42(13.4)	&129.8(14)\\
\hline
19	&1960-1964 	&149	&197	&337074	&201.3&118.08(41)	&136.22(32)\\
\hline
20	&1972-1976	&247	&325	&362000	&110.6&140.9(27)	&130.9(18)\\
\hline
21	&1982-1986	&229	&254	&396692	&164.5&131.97(19.7)	&139.3(15)\\
\hline
22	&1992-1996	&245	&289	&357607	&158.5&133.63(15.6)	&151.2(4.6)\\
\hline
23	&2004-2008	&509	&573	&267244	&120.8&132.16(9.4)	&137.8(14)\\
\hline
24	&2015-2019	&604	&913	&256737	&110.44&107.87(2.3)	&107.17(2.9)\\
\hline
25\textsuperscript{a}& & & & & &99.13 ± 14.97	&104.23± 17.35\\
\hline

\end{tabular}

Note: \textsuperscript{a} -  Represent the value calculated by both precursors: SLD and aa index

\end{table}

Shown in Figure $2$  is the 13 month smoothed  maximum  annual mean  sunspots number ($R_{max}$) of (n+1)$\textsuperscript{th}$ cycle plotted as a function of observed number of total SLD during the descending phase $( \sum SLD\textsubscript{dsc} $ ) of the n \textsuperscript{th} cycle. The best $\chi^2$ linear fit to the data shows correlation coefficient of 0.68, which led us to derive an empirical relation between these two indices for cycles $11$ - $24$  as follows.

\begin{equation}\label{eq1}
  { (R_{max})\textsubscript{n+1} = 154.70-0.092*(\sum SLD\textsubscript{n}) \textsubscript{dsc}  \pm 14.97 }
\end{equation}

where, $(R\textsubscript{max})\textsubscript{n+1}$ is the predicted 13 month smoothed maximum annual mean sunspot number of the (n+1)$\textsuperscript{th}$ cycle.

From figure 2, it is evident that point \#18 (corresponding to  R\textsubscript{max} of cycle 19 and SLDs in the descending phase of cycle 18) is an obvious outlier and thus been excluded in obtaining the empirical relation (1). Similar behavior of point \#18 is also noted in the figure 3, in which the correlation of  R\textsubscript{max} of preceding cycle is obtained against integrated aa index. This may be attributed to  either an underestimation of SLD and aa index, or an unusually  high magnetic field activity in the cycle 19 ( R\textsubscript{max}=201) compared to all of the investigated cycles where the maximum value of  R\textsubscript{max} has reached only to a level of 164.5.  The reason for this anomalously activity exhibited by predictors of cycle 18 has been investigated in detail by \citep{Wang2009}. They attributed an underestimation of the aa index (and radial IMF strength which is derived from the aa index) to cause this unusual behavior. Based on higher polar field strengths as estimated by \citet{sheeley2008century}, and a low-level of sunspot activity in 1954 (end of cycle 18), they have further argued that radial IMF strength and thus, aa index has been underestimated. This led to the exclusion of point \#18 from the correlation trend. In another investigation by \citet{podladchikova2017sunspot}, who established a correlation between the maximum amplitude of the solar cycle and the temporal behaviors of the sunspot number during the declining phase of the preceding cycle  in the form of ‘‘integral activity’’ which is the area under the sunspot cycle curve. After studying solar cycle 10-23, they found that a slow declining trend of the sunspot number, which exhibited a sudden increase during the year 1950-1954 (i.e. descending phase of cycle 18). Since the rate of declination of sunspots has decreased in this duration, they attributed this to   cause higher value  R\textsubscript{max} for cycle 19.This unusual behavior of cycle 19 led to its exclusion from further investigations by other investigators  \citet{Petrovay2010} and \citet{Clilverd2006}.

The total number 0f observed  SLD in the descending phase of cycle 24 are $( \sum SLD\textsubscript{24})\textsubscript{dsc} $ = 604, which yields $(R\textsubscript{max})\textsubscript{25}$ =99.13 $\pm$ 14.97 using equation~\ref{eq1}.
 This indicates that the amplitude of cycle $25$ will be  lower than the cycle $24$ ($R\textsubscript{max}$)\textsubscript{24} =$110.44$.

\begin{figure}
   \centering
   \includegraphics[width=0.6\textwidth, angle=0]{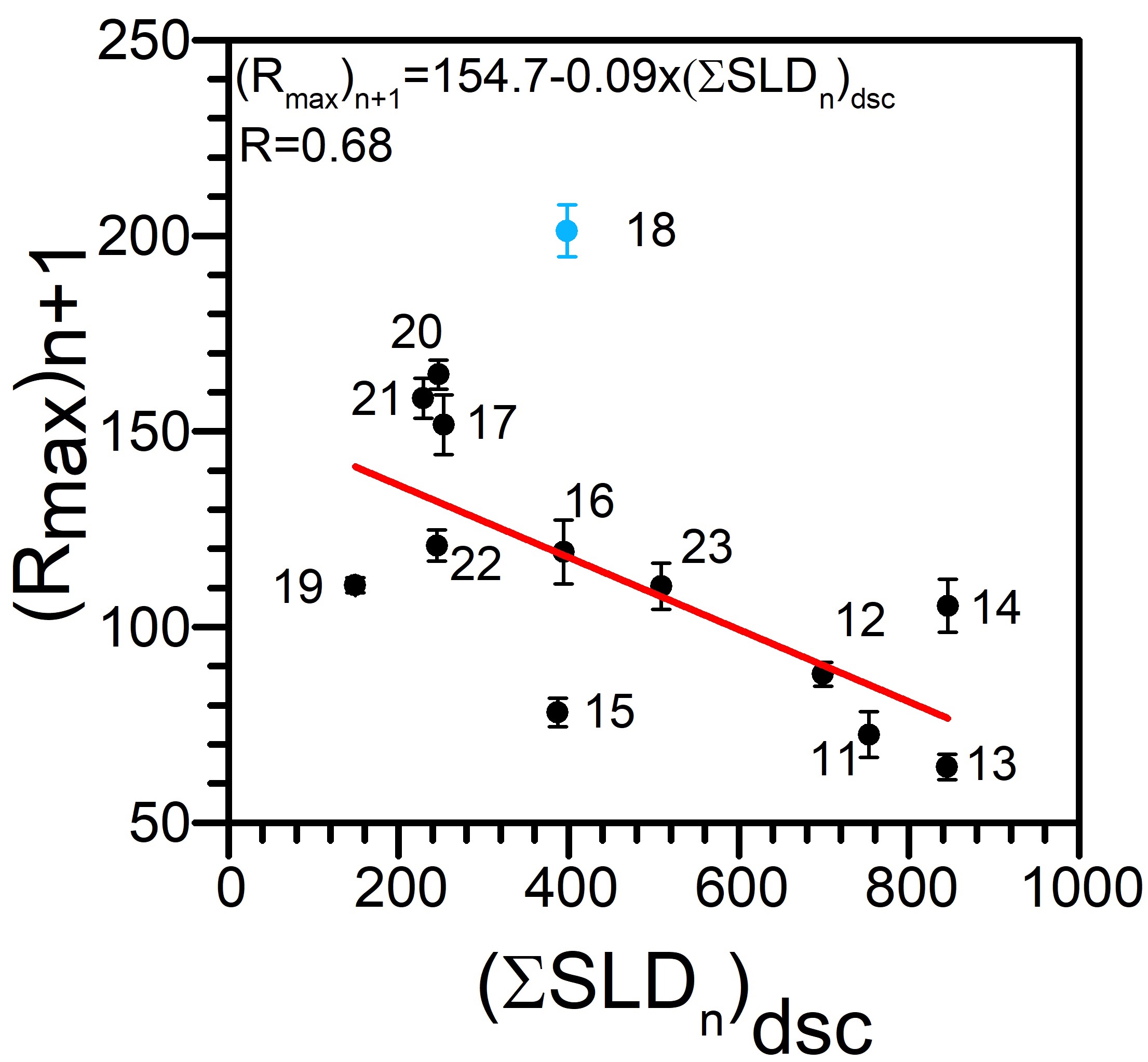}
\caption{The 13 month smoothed maximum annual mean sunspot number (R$_{max}$) of (n+1)$^{th}$ cycle plotted as a function of ($\sum$SLD)\textsubscript{dsc} of the n\textsuperscript{th} cycle. The data point corresponding to sum of the SLDs in the desending phase of  cycle 18 $( \sum SLD)\textsubscript{18}$ and  maximum amplitude of sunspot number (R\textsubscript{max}) of cycle 19, marked with point \#18 has been excluded from fitting equation. }
\label{fig2}
\end{figure}

Further, aa index during the descending phase of a given cycle has been found to be a good predictor of the amplitude of the following cycle \citep{Ohl1966, Jain1997, Lantos1998, Hathaway1999, Hathaway2009, Wang2009, Bhatt2009}. In order to apply the prediction scheme similar to that employed in the previous section on SLDs, we calculate the total value of aa index $ (\sum aa)\textsubscript{dsc}$ over the descending phase of the given n\textsuperscript{th} cycle. We provide the result in Table $3$ for the solar cycles  $11$ to $24$.

\begin{figure}
   \centering
   \includegraphics[width=0.6\textwidth, angle=0]{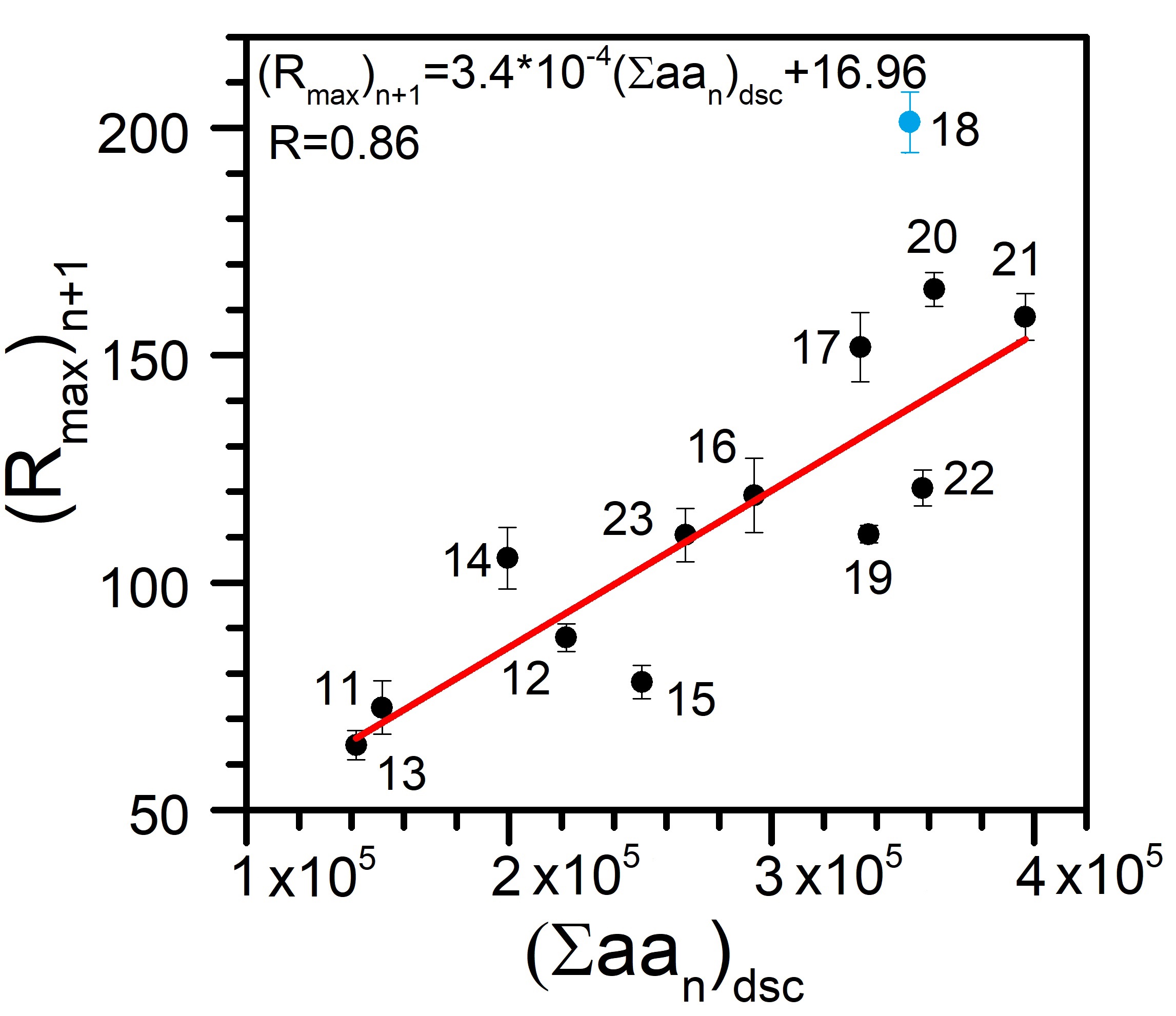}
\caption{The 13 month smoothed maximum annual mean sunspot numbers $(R_{max})$ of (n+1)$^{th}$ cycle plotted as a function of ($\sum$aa)\textsubscript{dsc} of the $n^{th}$ cycle.The correlation between them is (0.86). The data point corresponding to sum of the aa index  in the desending phase of  cycle 18 $( \sum aa)\textsubscript{18}$ and maximum amplitude of sunspot number (R\textsubscript{max}) of cycle 19, marked with point \#18 has been excluded from fitting equation.}
\label{fig3}
\end{figure}

We show in Figure 3 the relation between $13$ month smoothed  maximum annual mean sunspot numbers $(R\textsubscript{max})$ of (n+1)$\textsuperscript{th}$ cycle and  $( \sum aa)\textsubscript{dsc}$ of the n$\textsuperscript{th}$ cycle. The linear fit with a correlation coefficient  ($0.86$) is found with following empirical relation.

\begin{equation}\label{eq2}
  (R_{max})^{n+1} =3.4 *10^{-4}* (\sum aa\textsubscript{n})\textsubscript{dsc}+16.96) \pm 17.35
\end{equation}

where, $(R\textsubscript{max})\textsubscript{n+1}$ is the predicted $13$ month smoothed maximum annual mean sunspot numbers for the (n+1)$\textsuperscript{th}$ cycle.

Considering the cycle $24$ to have  ended in December $2019$, reference: \url{(www.sidc.be/silso/datafiles/node/166)}  we obtain $( \sum aa\textsubscript{dsc}) $= $256737$. Using this value in the relation ($2$) we obtain $(R\textsubscript{max})\textsubscript{25}$ = 104.23 $\pm$  17.35. This value of R\textsubscript{max}, derived from the aa index as the predictor agrees well with  that derived using SLD as a predictor (99.13 $\pm$  14.97). Therefore, this work identifies the potential of SLD as the precursor of next solar cycle.  Further, we have been able to confirm that the cycle $25$ will be weaker than the cycle $24$.

The table $3$ gives a comparison between the observed and calculated values of the $13$ month smoothed maximum annual mean sunspot numbers for both the  precursor methods. The comparative analysis indicates that in some cases, the calculated values from relation ~\ref{eq1}  agree well (relative to that calculated from relation ~\ref{eq2}) with the observed values of R\textsubscript{max} (for cycle $23$ and $24$) whereas vice-versa for others. Therefore, we find that, in general,the calculated values of R\textsubscript{max} from both the relations agree well with the observed values.

\subsection{Prediction of the Maximum Period of Solar Cycle 25}
Following Waldmeier effect \citep{Waldmeier1935} and our findings that the solar cycle $25$ will be weaker it is clear that the ascending phase of cycle $25$ is expected to be longer in duration. Thus it is important to predict the ascending period of cycle $25$. For this purpose we explore the relationship between the ascending phase period in years, $(P\textsubscript{asc})\textsubscript{n}$  and $(R\textsubscript{max})\textsubscript{n}$ . Shown in Figure 4 is the plot between $(P_{asc})_n$ and $(R_{max})_n$ for the cycles $11$-$24$. The best fit line shows a correlation value of  $-0.78$. The trend reveals that $(P\textsubscript{asc})\textsubscript{n}$ of a solar cycle decreases with the increase in $(R\textsubscript{max})\textsubscript{n}$, representing manifestation of the ``Waldmeier effect". The linear fit is expressed in the form of the following empirical relation:

\begin{equation}\label{eq3}
 (P\_{asc})_n = 5.7-0.015×(R\textsubscript{max})\textsubscript{n} \pm 0.43
\end{equation}

On substituting the value of $(R\textsubscript{max})\textsubscript{25}$ = 99.13 $\pm$  14.97 , derived from the SLD technique, we obtain $(P\textsubscript{asc})\textsubscript{25}$ = 4.21 $\pm$ 0.43 years. On the other hand, by substituting $(R\textsubscript{max})\textsubscript{25}$ =  104.23$\pm$ 17.35,  obtained from the aa method, we estimate $(P\textsubscript{asc})\textsubscript{n}$ = 4.14 $\pm$ 0.43 years. Results from both techniques reveal that ascending period for cycle $25$ is likely to be around 4.17 $\pm$ 0.43 years, ( average of both techniques), which suggests the peak amplitude of the cycle $25$ is likely to occur between February and March $2024$.

\begin{figure}
   \centering
   \includegraphics[width=0.6\textwidth, angle=0]{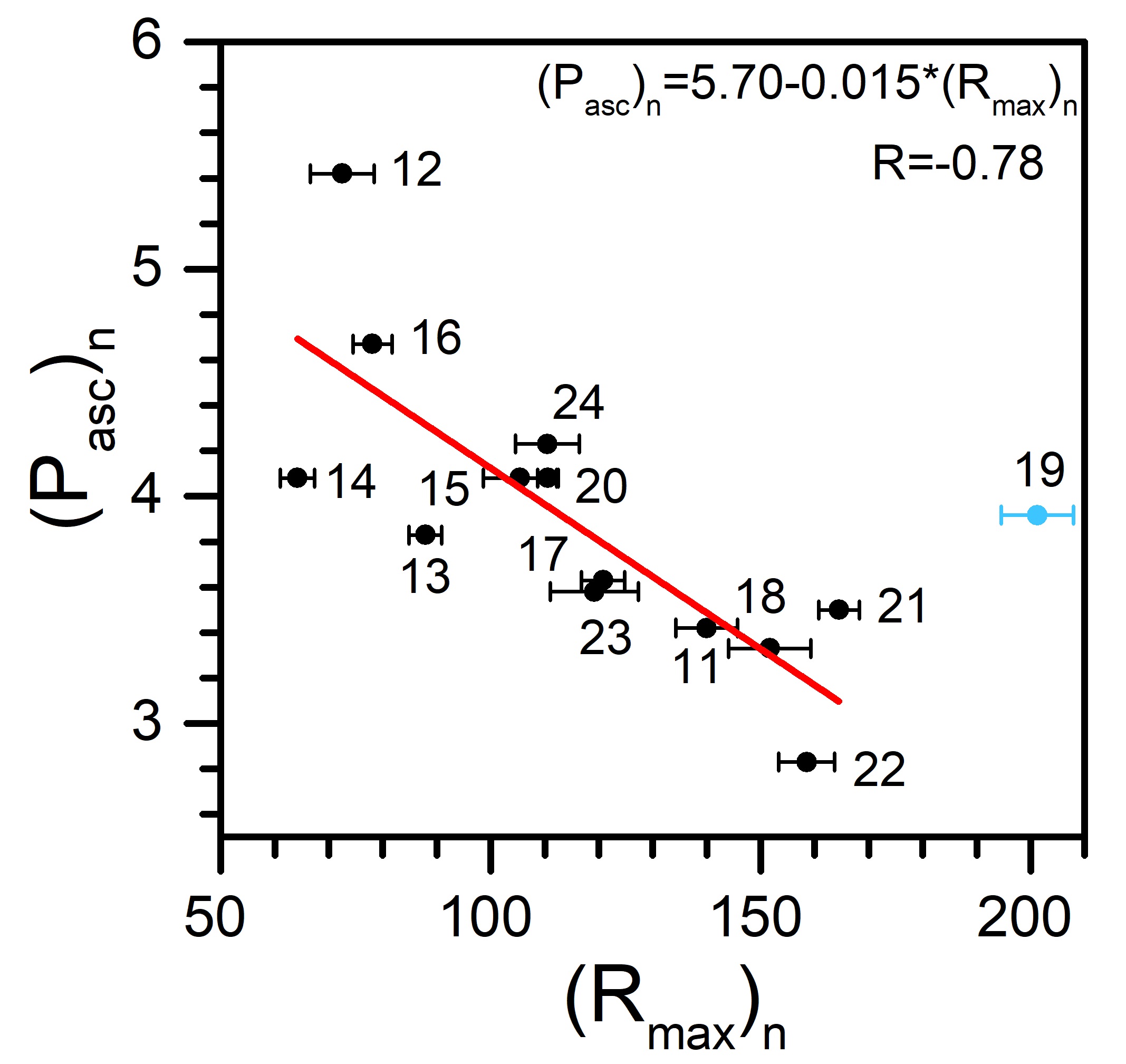}
\caption{The plot shows the ascending phase of the n$\textsuperscript{th}$ cycle as a function of 13 month smoothed maximum annual mean sunspot numbers (R$\textsubscript{max}$) of the same cycle. The data point representing the cycle 19 (marked by \#19) has been excluded from fitting equation.}
\label{fig4}
\end{figure}

\section{Discussion}
\label{sect:discussion}
In the current investigation we use the spotless days (SLDs) during the descending phase of a given cycle to predict the sunspot amplitude of the next cycle. Based on this technique we obtained maximum amplitude $(R\textsubscript{max})$ for cycle $25$ to be 99.13 $\pm$ 14.97. In addition to SLD as a predictor of the sunspot cycle characteristics, we also used sum of aa index as another solar activity proxy and found amplitude to be 104.23 $\pm$ 17.35. This is in good agreement with the amplitude estimated using SLD. Our predicted amplitude of sunspot number 99.13 ± 14.97  for cycle 25 is in agreement with the recent predictions made by few other investigators (within the range of 82-140); \citep {Upton2018, Jiang2018, Bhowmik2018, Petrovay2018, Bisoi2020, Sello2019, Dani2019, Covas2019, Labonville2019, Miao2020}. On the other hand, some studies \citep{Sarp2018, Li2018} contrarily predict the cycle 25 to have a higher value (150-170). This difference may be attributed to different predictors as well as numerical scheme used (such as non-linear prediction algorithm (154) and bimodal distribution (168)). Thus, we find the amplitude of solar cycle $25$ to be weaker than the current cycle $24$. Further, the period of the peak amplitude of cycle $25$ is predicted to be occurring between February and March $2024$.

\begin{table}
\centering
\caption{Prediction for cycle $25$}
\label{tab 4}
\vspace{0.5cm}
\begin{tabular}{|p{4cm}|p{3cm}|p{3cm}|}
\hline
\textbf{(Author year)}	&\textbf{Method} 	&\textbf{Maximum amplitude}\\
\hline
\citep{javaraiah2017will} 	& Cosine function+ linear fit	&30 - 40\\
\hline
\citep{Javaraiah2015}	&Maximum entropy method (MEM), and Morlet wavelet analysis.	&50±10\\
\hline
\citep{Covas2019} 	&Neural networks Spatiotemporal	&57 ±17\\
\hline
\citep{Kakad2017}	&Shannon entropy model	&63±11.3\\
\hline
\citep{Labonville2019}	&Solar dynamo models	&89(+/-29/14)\\
\hline
\citep{Attia2013}	&Neural networks Neuro-fuzzy	&90.7 ± 8\\
\hline
\citep{Gopalswamy2018}	&Microwave imaging observations	&89 and 59\\
\hline
\citep{Singh2019}	&Extrapolation method	&89 ± 9 and 78 ± 7\\
\hline
\citep{kitiashvili2016data}	&Data assimilation method	&90 ± 15\\
\hline
\citep{Jiang2018a}	 &SFT model	&93 to 125\\
\hline
\citep{chowdhury2021prediction}	&Precursor method Ap index 	&100.21 ± 15.06\\
\hline
\citep{Singh2017a}	&Hurst exponent	&103 ± 25\\
\hline
\citep{Hiremath2006}	&Forced and damped harmonic oscillator model	&110 (± 11)\\
\hline
\citep{Pishkalo2019}	&Polar field as precursor	&116±12\\
\hline
\citep{Hawkes2018a}	&Magnetic helicity	&117\\
\hline
\citep{Helal2013a}	&Precursor technique of spotless	&118.2\\
\hline
\citep{Pesnell2018}	&Employing solar dynamo index	&135 ±25\\
\hline
\citep{Du2020a}	&Logarithmic relationship between Rm and aamin	&151.1±16.9\\
\hline
\citep{Sarp2018}	&Non-linear prediction algorithm	&154± 12\\
\hline
\end{tabular}
\end{table}

We have also shown in Table $2$  that  numbers of SLD are continuously increasing  while sum of aa index and sunspot numbers are decreasing from cycle 21 . A total 509 SLD were observed during the  descending phase of solar cycle $23$, significantly higher than the value of $229$ and $245$ for  solar cycles $21$ and $22$ respectively. On the other hand, in the solar cycle $24$, an extremely low sunspot activity during ascending ($2009$-$2012$) as well as descending phase ($2015-2019$) have been reported, and 604 SLDs have been observed. The continuous increases of  numbers of SLD and sum of the aa index from cycle 21 and inferences from the current investigation reveal that the cycle $24$ and $25$ are likely to be the part of the series of $2$-$4$ successive weak and long cycles. This may suggest that the sun  has been progressing towards  or in the Gleissberg minimum or grand solar minimum  \citep{Petrovay2018, Yousef2003, feynman1990}.
 Considering low amplitude of cycle $25$ and decreasing trend of solar surface magnetic field,  it is likely that cycle $26$ may also be a weak cycle and if this trend continues then it appears that  Sun may be heading towards next Grand Minimum. However, while our sunspot number predictions are in agreement with that obtained by several investigators in the past (see Table 4), they do not necessarily concur on our inference on Sun to be heading towards grand solar minimum. For example,\citep{Bhowmik2018}  predicted the maximum amplitude of the 25\textsuperscript{th} cycle to be 118 with a peak in 2024,which is  in agreement with our prediction. However, their ensemble forecast model indicates the reversal of weakening trend of solar activity.

	   It may be noted that the last Maunder Minimum was during $1645$-$1715$ AD, and the end of upcoming cycle $26$ would be nearly in $2043$ AD, almost $400$ years after the Maunder Minimum. Our above conclusions is further strengthened from the investigation carried out by \citep{Cameron2016} who  showed that the measurements of the dipole moment of about 2G at the end of cycle $23$,  is close to simulated value  of  about 2.5 $\pm$ 1.1 G for cycle $25$ in $2020$. Based on the best correlation between the dipole moment during the solar minimum and the strength of the next solar cycle, \citep{Cameron2016} further suggested that cycle $25$ will be of moderate amplitude, not much higher than that of the current cycle. It is widely known from several evidences that the strength of the sun's polar fields near the time of a sunspot cycle minimum determines the strength of the following solar activity cycle. \citep{Hathaway2016} applied their Advective Flux Transport code, with flows well constrained by observations, to simulate the evolution of the sun's polar magnetic fields from early $2016$ to the end of $2020$, near the expected time of cycle $24/25$ minimum. Considering various limitations and uncertainties in simulations they find that the average strength of the polar fields near the end of cycle $24$ will be similar to that measured near the end of cycle $23$, indicating that cycle $25$ will be similar in strength to the current cycle $24$. Further, \citep{MacarioRojas2018} also showed that the cycle $25$ will be a weak solar cycle with slow rise time (-14.4 per cent) and maximum activity  ($\pm$19.5 per cent) with respect to the current cycle $24$. Table $4$ summarizes the various predictions of several investigators for the amplitude of cycle $25$. The table contains the author as well as year of publication, prediction methods and predicted values in column $1$, $2$, and  $3$  respectively. From the table, it is clear that our results  concur with the most of previous investigators in the sense that the solar cycle $25$ has amplitude lower than the amplitude of solar cycle $24$. On the other hand, there are few authors who claim that the cycle $25$ may be of similar or higher amplitude than the cycle $24$. Moreover, our results from both techniques (SLD and aa) fall within the uncertainty of the NOAA/NASA consensus predictions on December $9$, $2019$ that strength of upcoming solar cycle $25$ would be comparable to cycle $24$ (https:// www. swpc. noaa. Gov/ news /solar-cycle-25-forecast-update).

\section{Conclusions}
\label{sect:conclusion}
We have applied the available data for the SLD from cycle $11$ to $24$ ($1874$-$2019$). The occurrence frequency of SLD shows negative exponential  distribution with respect to the consecutive days. A new prediction technique is proposed based on the number of SLD observed in the declined phase of solar cycle which shows that the maximum of the $13$ month smoothed maximum annual mean  sunspot number for cycle $25$ will be 99.13 $\pm$ 14.97. The  SLD occurrence density per cycle is found to be increase from cycle $21$ while  sum of aa index and 13-month smoothed  monthly sunspot number are decreasing during the descending phase of a cycle in particular (cf. table $2$). This trend is  indicative of weakening in solar surface magnetic field  of the sun. In view of of our analysis which resulted in the fact that solar cycle $25$ is likely to be weaker in strength relative to the current cycle $24$, we conclude that the sun may be proceeding towards next grand minimum.

\begin{acknowledgements}
Authors received the data for Sunspot Index and Long-term Solar Observations (SILSO) site of SIDC, WDC, Royal Observatory, Belgium. Cooperation received from President, KSV is sincerely acknowledged. RJ was invited by GSFC/ NASA to work on this investigation. AKA acknowledges the funding support from NSFC-11950410498 and KLSA-202010 grants. We earnestly thanks the referee for valuable remarks which has improved the paper.
\end{acknowledgements}

\bibliographystyle{aasjournal_2019}
\bibliography{bibtex}

\begin{thebibliography}{}
\expandafter\ifx\csname natexlab\endcsname\relax\def\natexlab#1{#1}\fi
\providecommand{\url}[1]{\href{#1}{#1}}
\providecommand{\dodoi}[1]{doi:~\href{http://doi.org/#1}{\nolinkurl{#1}}}
\providecommand{\doeprint}[1]{\href{http://ascl.net/#1}{\nolinkurl{http://ascl.net/#1}}}
\providecommand{\doarXiv}[1]{\href{https://arxiv.org/abs/#1}{\nolinkurl{https://arxiv.org/abs/#1}}}

\bibitem[{{Ata{\c{c}}} {et~al.}(2006){Ata{\c{c}}}, {{\"O}zg{\"u}{\c{c}}}, \&
  {Rybak}}]{Atac2006}
{Ata{\c{c}}}, T., {{\"O}zg{\"u}{\c{c}}}, A., \& {Rybak}, J. 2006, \solphys,
  237, 433, \dodoi{10.1007/s11207-006-0017-5}

\bibitem[{{Attia} {et~al.}(2013){Attia}, {Ismail}, \& {Basurah}}]{Attia2013}
{Attia}, A.-F., {Ismail}, H.~A., \& {Basurah}, H.~M. 2013, \apss, 344, 5,
  \dodoi{10.1007/s10509-012-1300-6}

\bibitem[{{Bhatt} {et~al.}(2009){Bhatt}, {Jain}, \& {Aggarwal}}]{Bhatt2009}
{Bhatt}, N.~J., {Jain}, R., \& {Aggarwal}, M. 2009, \solphys, 260, 225,
  \dodoi{10.1007/s11207-009-9439-1}

\bibitem[{{Bhowmik} \& {Nandy}(2018)}]{Bhowmik2018}
{Bhowmik}, P., \& {Nandy}, D. 2018, Nature Communications, 9, 5209,
  \dodoi{10.1038/s41467-018-07690-0}

\bibitem[{{Bisoi} {et~al.}(2020){Bisoi}, {Janardhan}, \&
  {Ananthakrishnan}}]{Bisoi2020}
{Bisoi}, S.~K., {Janardhan}, P., \& {Ananthakrishnan}, S. 2020, Journal of
  Geophysical Research (Space Physics), 125, e27508,
  \dodoi{10.1029/2019JA027508}

\bibitem[{{Braun} {et~al.}(2005){Braun}, {Christl}, {Rahmstorf}, {Ganopolski},
  {Mangini}, {Kubatzki}, {Roth}, \& {Kromer}}]{Braun2005}
{Braun}, H., {Christl}, M., {Rahmstorf}, S., {et~al.} 2005, \nat, 438, 208,
  \dodoi{10.1038/nature04121}

\bibitem[{{Cameron} {et~al.}(2016){Cameron}, {Jiang}, \&
  {Sch{\"u}ssler}}]{Cameron2016}
{Cameron}, R.~H., {Jiang}, J., \& {Sch{\"u}ssler}, M. 2016, \apjl, 823, L22,
  \dodoi{10.3847/2041-8205/823/2/L22}

\bibitem[{{Carrasco} {et~al.}(2016){Carrasco}, {Aparicio}, {Vaquero}, \&
  {Gallego}}]{Carrasco2016}
{Carrasco}, V.~M.~S., {Aparicio}, A.~J.~P., {Vaquero}, J.~M., \& {Gallego},
  M.~C. 2016, \solphys, 291, 3045, \dodoi{10.1007/s11207-016-0998-7}

\bibitem[{Chowdhury {et~al.}(2021)Chowdhury, Jain, Ray, Burud, \&
  Chakrabarti}]{chowdhury2021prediction}
Chowdhury, P., Jain, R., Ray, P., Burud, D., \& Chakrabarti, A. 2021, Solar
  Physics, 296, 1

\bibitem[{{Clilverd} {et~al.}(2006){Clilverd}, {Clarke}, {Ulich}, {Rishbeth},
  \& {Jarvis}}]{Clilverd2006}
{Clilverd}, M.~A., {Clarke}, E., {Ulich}, T., {Rishbeth}, H., \& {Jarvis},
  M.~J. 2006, Space Weather, 4, S09005, \dodoi{10.1029/2005SW000207}

\bibitem[{{Covas} {et~al.}(2019){Covas}, {Peixinho}, \&
  {Fernandes}}]{Covas2019}
{Covas}, E., {Peixinho}, N., \& {Fernandes}, J. 2019, \solphys, 294, 24,
  \dodoi{10.1007/s11207-019-1412-z}

\bibitem[{{Dani} \& {Sulistiani}(2019)}]{Dani2019}
{Dani}, T., \& {Sulistiani}, S. 2019, in Journal of Physics Conference Series,
  Vol. 1231, Journal of Physics Conference Series, 012022,
  \dodoi{10.1088/1742-6596/1231/1/012022}

\bibitem[{{Du}(2020)}]{Du2020a}
{Du}, Z.~L. 2020, \apss, 365, 104, \dodoi{10.1007/s10509-020-03818-1}

\bibitem[{{El-Borie} {et~al.}(2020){El-Borie}, {El-Taher}, {Thabet}, {Ibrahim},
  {Aly}, \& {Bishara}}]{ElBorie2020}
{El-Borie}, M.~A., {El-Taher}, A.~M., {Thabet}, A.~A., {et~al.} 2020, \apj,
  898, 73, \dodoi{10.3847/1538-4357/ab9d21}

\bibitem[{Feynman \& Gabriel(1990)}]{feynman1990}
Feynman, J., \& Gabriel, S.~B. 1990, Solar Physics, 127, 393

\bibitem[{{Gopalswamy} {et~al.}(2018){Gopalswamy}, {M{\"a}kel{\"a}}, {Yashiro},
  \& {Akiyama}}]{Gopalswamy2018}
{Gopalswamy}, N., {M{\"a}kel{\"a}}, P., {Yashiro}, S., \& {Akiyama}, S. 2018,
  Journal of Atmospheric and Solar-Terrestrial Physics, 176, 26,
  \dodoi{10.1016/j.jastp.2018.04.005}

\bibitem[{{Hamid} \& {Galal}(2006)}]{Hamid2006}
{Hamid}, R.~H., \& {Galal}, A.~A. 2006, in Solar Activity and its Magnetic
  Origin, ed. V.~{Bothmer} \& A.~A. {Hady}, Vol. 233, 413--416,
  \dodoi{10.1017/S1743921306002390}

\bibitem[{{Harvey}(1992)}]{Harvey1992}
{Harvey}, K.~L. 1992, in Astronomical Society of the Pacific Conference Series,
  Vol.~27, The Solar Cycle, ed. K.~L. {Harvey}, 335

\bibitem[{{Harvey} \& {White}(1999)}]{Harvey1999}
{Harvey}, K.~L., \& {White}, O.~R. 1999, \jgr, 104, 19759,
  \dodoi{10.1029/1999JA900211}

\bibitem[{{Hathaway}(2009)}]{Hathaway2009}
{Hathaway}, D.~H. 2009, {Solar Cycle Forecasting}, Vol.~32, 401,
  \dodoi{10.1007/978-1-4419-0239-9_20}

\bibitem[{{Hathaway} \& {Upton}(2016)}]{Hathaway2016}
{Hathaway}, D.~H., \& {Upton}, L.~A. 2016, Journal of Geophysical Research
  (Space Physics), 121, 10,744, \dodoi{10.1002/2016JA023190}

\bibitem[{{Hathaway} {et~al.}(1999){Hathaway}, {Wilson}, \&
  {Reichmann}}]{Hathaway1999}
{Hathaway}, D.~H., {Wilson}, R.~M., \& {Reichmann}, E.~J. 1999, \jgr, 104,
  22375, \dodoi{10.1029/1999JA900313}

\bibitem[{{Hathaway} {et~al.}(2002){Hathaway}, {Wilson}, \&
  {Reichmann}}]{Hathaway2002}
---. 2002, \solphys, 211, 357, \dodoi{10.1023/A:1022425402664}

\bibitem[{{Hawkes} \& {Berger}(2018)}]{Hawkes2018a}
{Hawkes}, G., \& {Berger}, M.~A. 2018, \solphys, 293, 109,
  \dodoi{10.1007/s11207-018-1332-3}

\bibitem[{{Helal} \& {Galal}(2013{\natexlab{a}})}]{Helal2013}
{Helal}, H.~R., \& {Galal}, A.~A. 2013{\natexlab{a}}, Journal of Advanced
  Research, 4, 275, \dodoi{10.1016/j.jare.2012.10.002}

\bibitem[{{Helal} \& {Galal}(2013{\natexlab{b}})}]{Helal2013a}
---. 2013{\natexlab{b}}, Journal of Advanced Research, 4, 275,
  \dodoi{10.1016/j.jare.2012.10.002}

\bibitem[{{Hiremath}(2006)}]{Hiremath2006}
{Hiremath}, K.~M. 2006, \aap, 452, 591, \dodoi{10.1051/0004-6361:20042619}

\bibitem[{{Jain}(1986)}]{Jain1986}
{Jain}, R. 1986, \mnras, 223, 877, \dodoi{10.1093/mnras/223.4.877}

\bibitem[{{Jain}(1997)}]{Jain1997}
---. 1997, \solphys, 176, 431, \dodoi{10.1023/A:1004973827442}

\bibitem[{{Javaraiah}(2015)}]{Javaraiah2015}
{Javaraiah}, J. 2015, \na, 34, 54, \dodoi{10.1016/j.newast.2014.04.001}

\bibitem[{Javaraiah(2017)}]{javaraiah2017will}
Javaraiah, J. 2017, Solar Physics, 292, 1

\bibitem[{{Javaraiah}(2019)}]{Javaraiah2019}
{Javaraiah}, J. 2019, \solphys, 294, 64, \dodoi{10.1007/s11207-019-1442-6}

\bibitem[{{Jiang} \& {Cao}(2018)}]{Jiang2018a}
{Jiang}, J., \& {Cao}, J. 2018, Journal of Atmospheric and Solar-Terrestrial
  Physics, 176, 34, \dodoi{10.1016/j.jastp.2017.06.019}

\bibitem[{{Jiang} {et~al.}(2018){Jiang}, {Wang}, {Jiao}, \& {Cao}}]{Jiang2018}
{Jiang}, J., {Wang}, J.-X., {Jiao}, Q.-R., \& {Cao}, J.-B. 2018, \apj, 863,
  159, \dodoi{10.3847/1538-4357/aad197}

\bibitem[{{Kakad} {et~al.}(2017){Kakad}, {Kakad}, \& {Ramesh}}]{Kakad2017}
{Kakad}, B., {Kakad}, A., \& {Ramesh}, D.~S. 2017, \solphys, 292, 181,
  \dodoi{10.1007/s11207-017-1207-z}

\bibitem[{{Kilcik} {et~al.}(2010){Kilcik}, {{\"O}zg{\"u}{\c{c}}}, {Rozelot}, \&
  {Ata{\c{c}}}}]{Kilcik2010}
{Kilcik}, A., {{\"O}zg{\"u}{\c{c}}}, A., {Rozelot}, J.~P., \& {Ata{\c{c}}}, T.
  2010, \solphys, 264, 255, \dodoi{10.1007/s11207-010-9567-7}

\bibitem[{{Kilcik} {et~al.}(2012){Kilcik}, {Yurchyshyn}, {Rempel}, {Abramenko},
  {Kitai}, {Goode}, {Cao}, \& {Watanabe}}]{Kilcik2012}
{Kilcik}, A., {Yurchyshyn}, V.~B., {Rempel}, M., {et~al.} 2012, \apj, 745, 163,
  \dodoi{10.1088/0004-637X/745/2/163}

\bibitem[{Kitiashvili(2016)}]{kitiashvili2016data}
Kitiashvili, I.~N. 2016, The Astrophysical Journal, 831, 15

\bibitem[{{Krivova} \& {Solanki}(2002)}]{Krivova2002}
{Krivova}, N.~A., \& {Solanki}, S.~K. 2002, \aap, 394, 701,
  \dodoi{10.1051/0004-6361:20021063}

\bibitem[{{Labonville} {et~al.}(2019){Labonville}, {Charbonneau}, \&
  {Lemerle}}]{Labonville2019}
{Labonville}, F., {Charbonneau}, P., \& {Lemerle}, A. 2019, \solphys, 294, 82,
  \dodoi{10.1007/s11207-019-1480-0}

\bibitem[{{Lantos} \& {Richard}(1998)}]{Lantos1998}
{Lantos}, P., \& {Richard}, O. 1998, \solphys, 182, 231,
  \dodoi{10.1023/A:1005087612053}

\bibitem[{{Li} {et~al.}(2018){Li}, {Kong}, {Xie}, {Xiang}, \& {Xu}}]{Li2018}
{Li}, F.~Y., {Kong}, D.~F., {Xie}, J.~L., {Xiang}, N.~B., \& {Xu}, J.~C. 2018,
  Journal of Atmospheric and Solar-Terrestrial Physics, 181, 110,
  \dodoi{10.1016/j.jastp.2018.10.014}

\bibitem[{{Li} {et~al.}(2005){Li}, {Gao}, \& {Su}}]{Li2005}
{Li}, K.-J., {Gao}, P.-X., \& {Su}, T.-W. 2005, \cjaa, 5, 539,
  \dodoi{10.1088/1009-9271/5/5/011}

\bibitem[{{Macario-Rojas} {et~al.}(2018){Macario-Rojas}, {Smith}, \&
  {Roberts}}]{MacarioRojas2018}
{Macario-Rojas}, A., {Smith}, K.~L., \& {Roberts}, P. C.~E. 2018, \mnras, 479,
  3791, \dodoi{10.1093/mnras/sty1625}

\bibitem[{{McKinnon} \& {Waldmeier}(1987)}]{McKinnon1987}
{McKinnon}, J.~A., \& {Waldmeier}, M. 1987, {Sunspot numbers, 1610-1985 : based
  on ``The sunspot activity in the years 1610-1960''}

\bibitem[{{Miao} {et~al.}(2020){Miao}, {Wang}, {Ren}, \& {Li}}]{Miao2020}
{Miao}, J., {Wang}, X., {Ren}, T.-L., \& {Li}, Z.-T. 2020, Research in
  Astronomy and Astrophysics, 20, 004, \dodoi{10.1088/1674-4527/20/1/4}

\bibitem[{{Nandy} \& {Martens}(2007)}]{Nandy2007}
{Nandy}, D., \& {Martens}, P.~C.~H. 2007, Advances in Space Research, 40, 891,
  \dodoi{10.1016/j.asr.2007.01.079}

\bibitem[{Ohl(1966)}]{Ohl1966}
Ohl, A. 1966, Soln. Dannye, 12, 84

\bibitem[{{Pesnell} \& {Schatten}(2018)}]{Pesnell2018}
{Pesnell}, W.~D., \& {Schatten}, K.~H. 2018, \solphys, 293, 112,
  \dodoi{10.1007/s11207-018-1330-5}

\bibitem[{{Petrovay}(2010)}]{Petrovay2010}
{Petrovay}, K. 2010, Living Reviews in Solar Physics, 7, 6,
  \dodoi{10.12942/lrsp-2010-6}

\bibitem[{{Petrovay}(2020)}]{Petrovay2020}
---. 2020, Living Reviews in Solar Physics, 17, 2,
  \dodoi{10.1007/s41116-020-0022-z}

\bibitem[{{Petrovay} {et~al.}(2018){Petrovay}, {Nagy}, {Gerj{\'a}k}, \&
  {Juh{\'a}sz}}]{Petrovay2018}
{Petrovay}, K., {Nagy}, M., {Gerj{\'a}k}, T., \& {Juh{\'a}sz}, L. 2018, Journal
  of Atmospheric and Solar-Terrestrial Physics, 176, 15,
  \dodoi{10.1016/j.jastp.2017.12.011}

\bibitem[{{Pishkalo}(2019)}]{Pishkalo2019}
{Pishkalo}, M.~I. 2019, \solphys, 294, 137, \dodoi{10.1007/s11207-019-1520-9}

\bibitem[{Podladchikova {et~al.}(2017)Podladchikova, Van~der Linden, \&
  Veronig}]{podladchikova2017sunspot}
Podladchikova, T., Van~der Linden, R., \& Veronig, A.~M. 2017, The
  Astrophysical Journal, 850, 81

\bibitem[{{Sarp} {et~al.}(2018){Sarp}, {Kilcik}, {Yurchyshyn}, {Rozelot}, \&
  {Ozguc}}]{Sarp2018}
{Sarp}, V., {Kilcik}, A., {Yurchyshyn}, V., {Rozelot}, J.~P., \& {Ozguc}, A.
  2018, \mnras, 481, 2981, \dodoi{10.1093/mnras/sty2470}

\bibitem[{{Sello}(2019)}]{Sello2019}
{Sello}, S. 2019, arXiv e-prints, arXiv:1902.05294.
\newblock \doarXiv{1902.05294}

\bibitem[{Sheeley~Jr(2008)}]{sheeley2008century}
Sheeley~Jr, N. 2008, The Astrophysical Journal, 680, 1553

\bibitem[{{Singh} \& {Bhargawa}(2017)}]{Singh2017a}
{Singh}, A.~K., \& {Bhargawa}, A. 2017, \apss, 362, 199,
  \dodoi{10.1007/s10509-017-3180-2}

\bibitem[{{Singh} \& {Bhargawa}(2019)}]{Singh2019}
---. 2019, \apss, 364, 12, \dodoi{10.1007/s10509-019-3500-9}

\bibitem[{{Upton} \& {Hathaway}(2018)}]{Upton2018}
{Upton}, L.~A., \& {Hathaway}, D.~H. 2018, \grl, 45, 8091,
  \dodoi{10.1029/2018GL078387}

\bibitem[{{Usoskin} \& {Mursula}(2003)}]{Usoskin2003}
{Usoskin}, I.~G., \& {Mursula}, K. 2003, \solphys, 218, 319,
  \dodoi{10.1023/B:SOLA.0000013049.27106.07}

\bibitem[{{Waldmeier}(1935)}]{Waldmeier1935}
{Waldmeier}, M. 1935, Astronomische Mitteilungen der Eidgen\&ouml;ssischen
  Sternwarte Zurich, 14, 105

\bibitem[{{Waldmeier}(1961)}]{Waldmeier1961}
---. 1961, {The sunspot-activity in the years 1610-1960}

\bibitem[{{Wang} \& {Sheeley}(2009)}]{Wang2009}
{Wang}, Y.~M., \& {Sheeley}, N.~R. 2009, \apjl, 694, L11,
  \dodoi{10.1088/0004-637X/694/1/L11}

\bibitem[{{Wilson}(1995)}]{Wilson1995}
{Wilson}, R.~M. 1995, \solphys, 158, 197, \dodoi{10.1007/BF00680842}

\bibitem[{{Wilson} {et~al.}(1996){Wilson}, {Hathaway}, \&
  {Reichmann}}]{Wilson1996}
{Wilson}, R.~M., {Hathaway}, D.~H., \& {Reichmann}, E.~J. 1996, \jgr, 101,
  19967, \dodoi{10.1029/96JA01820}

\bibitem[{{Yousef}(2003)}]{Yousef2003}
{Yousef}, S. 2003, in ESA Special Publication, Vol. 535, Solar Variability as
  an Input to the Earth's Environment, ed. A.~{Wilson}, 177--180

\end{thebibliography}

\label{lastpage}

\end{document}